\documentclass{article}

\usepackage{PRIMEarxiv}
\usepackage{epsfig}
\usepackage[utf8]{inputenc} 
\usepackage[T1]{fontenc}    
\usepackage{hyperref}       
\usepackage{url} 
\usepackage{comment}
\usepackage{booktabs}       
\usepackage{amsfonts}       
\usepackage{nicefrac}       
\usepackage{microtype}      
\usepackage{lipsum}
\usepackage{fancyhdr}       
\usepackage{graphicx}       
\graphicspath{{media/}}     
\usepackage{algorithm}
\usepackage{algorithmic}
\usepackage[square,sort,comma,numbers]{natbib}
\usepackage{lipsum}	
\usepackage{multirow}
\usepackage{url}            
\usepackage{booktabs}
\usepackage{epsfig}
\usepackage{subcaption}
\usepackage{calc}
\usepackage{amssymb}
\usepackage{amstext}
\usepackage{amsmath}
\usepackage{amsthm}
\usepackage{tikz}
\usepackage{caption}
\usepackage{natbib}
\usepackage{amsmath}
\DeclareMathOperator{\lcm}{lcm}
\title{Cryptanalysis of a privacy-preserving behavior-oriented authentication scheme
\thanks{\textit{\underline{Citation}}: 
\textbf{Eskeland, S. and Baig, A. (2022). Cryptanalysis of a Privacy-preserving Behavior-oriented Authentication Scheme. In Proceedings of the 19th International Conference on Security and Cryptography - SECRYPT, ISBN 978-989-758-590-6; ISSN 2184-7711, pages 299-304. DOI: 10.5220/001114030000328
}} 
}

\author{
Sigurd Eskeland and Ahmed Fraz Baig\\
 Norwegian Computing Center \\ Postboks 114 Blindern \\ 0314 Oslo, Norway
  \texttt{\{sigurd,baig\}@nr.no} 
}

\begin{document}

\maketitle

\begin{abstract}
Continuous authentication has been proposed as a complementary security mechanism to password-based authentication for computer devices that are handled directly by humans, such as smart phones.
Continuous authentication has some privacy issues as certain user features and actions are revealed to the authentication server, which is not assumed to be trusted.
Wei et al. proposed in 2021 a privacy-preserving protocol for behavioral authentication that utilizes homomorphic encryption.
The encryption prevents the server from obtaining sampled user features.
In this paper, we show that the Wei et al. scheme is insecure regarding both an honest-but-curious server and an active eavesdropper.
We present two attacks:
The first attack enables the authentication server to obtain the secret user key, plaintext behavior template and plaintext authentication behavior data from encrypted data.
The second attack enables an active eavesdropper to restore the plaintext authentication behavior data from the transmitted encrypted data.
\end{abstract}

\keywords{Privacy-preserving behavioral authentication, Cryptographic protocols, Homomorphic encryption, Cryptanalysis, Continuous authentication}


\section{\uppercase{Introduction}}

Continuous authentication has been proposed as a complementary security measure for computer devices that are handled directly by humans, such as smart phones, in addition to common authentication methods, such as passwords, iris recogniztion, etc.
The supposed advantage is a passive and seamless authentication mechanism that does not require user attention, like re-typing of passwords.
While conventional authentication methods are session-oriented by which the device remains unlocked during the time period of the session, 
the idea of continuous authentication is that the authentication process is conducted at events of relevant user activity.
The time window of access is much smaller than for session-oriented approaches.
One purported benefit of continuous authentication over session-oriented approaches is that if a smart phone for a moment becomes  accessible to someone else while it is unlocked, the continuous authentication mechanism will supposedly not recognize the other person.
This will cause the authentication to fail and the phone will then lock.

Behavioral authentication is the most important category of modalities for continuous authentication.
The premise of behavioral authentication is that there is a uniqueness to the way that a person moves and acts, like walking style, typing style, or handling of devices, and recognizing such unique patterns is sufficient for identifying the person.
Behavioral modalities (or modes) include gait, screen touch (known as touch dynamics), and typing (keystroke dynamics).
Continuous authentication is realized by continuously monitoring and collecting user behavior data pertaining to a specific modality, and checking whether they are consistent with behavior reference template data collected during user enrollment.
In contrast, biometric authentication modalities such as face and iris recognition are oftentimes considered as continuous authentication as well.
However, since such modalities require some user attention, they are not entirely passive and seamless, and are therefore somewhat inconsistent the aspect of continuity.

It has been noted that continual monitoring and data collection of user activity can be considered invasive and that it causes privacy concerns, as it may reveal certain user actions and whereabouts while the user is in contact with the device.
Moreover, certain private user characteristics may be deductable, such as age group, gender, etc.
In conclusion, such authentication methods have indeed some privacy challenges.

Homomorphic encryption techniques have been suggested to mitigate the mentioned privacy challenges for continuous authentication modalities, as homomorphic encryption permits certain kinds of computations to be performed on encrypted data without first decrypting them.
This allows encrypted data to be outsourced to commercial cloud environments for processing, all while encrypted.


\citet{wei:2020} proposed in 2021 a privacy-preserving protocol for behavioral authentication, which assumes additive homomorphisms by building on the Paillier public key cryptosystem~\citep{Paillier:99}.
The authors claim that the scheme is secure with regard to both an honest-but-curious server and an active eavesdropper.
The eavesdropper is assumed to read and modify the communication between the user device and the authentication server.

In this paper, we show that the Wei et al. scheme is insecure regarding both an honest-but-curious server and an active eavesdropper.
We present two attacks, in which the first enables the authentication server to obtain the behavioral plaintext template, the authentication plaintext data, and the user's secret encryption key plaintext from the ciphertext data.
The second attack enables an eavesdropper to obtain authentication behavior plaintext data from the transmitted encrypted data.


\section{\uppercase{Related work}}

A few privacy-preserving schemes have been proposed for different types of modalities of behavior-based and context-based user authentication.
\citet{govindarajan:2013} proposed a privacy-preserving protocol for touch dynamics-based authentication.
Their scheme utilizes a private comparison protocol proposed by \citet{erkin:2009} and the homomorphic DGK encryption algorithm proposed by \citet{Damgaard:2008}.
Note that the \citet{erkin:2009} comparison protocol is based on the private comparison protocol proposed by~\citet{damgaard:2007,damgaard:2009}.
The scheme of Govindarajan et al. does not reveal anything, because it makes comparisons in the encrypted domain.

\citet{safa:2014} proposed a generic framework for privacy-preserving implicit authentication by utilizing context data, such as location data, device-specific data, wifi connection, browsing history, etc.
It utilizes homomorphic encryption and order-preserving encryption, and Average Absolute Deviation to compute the similarity between input and reference templates.

\citet{domingo:2015} proposed an privacy-preserving authentication scheme using context features.
It uses the Paillier cryptosystem and a private set intersection computation protocol proposed by the same authors~\citep{Justicia:2014}.
Set intersection is used to determine the dissimilarity between reference data and input data.

The privacy-preserving authentication scheme proposed by \citet{shahandashti:2015} assumes context features, and is based on order-preserving symmetric encryption (OPSE) and additive homomorphic encryption.
The cryptographic primitives are generic, but the authors suggest the OPSE scheme proposed by  \citet{Boldyreva:2009} and the Paillier public key scheme.

A potential problem with \citep{safa:2014,domingo:2015,shahandashti:2015} is that context-aware modes cannot differentiate if the user is present or not, such as if the device is stolen within the specified domain, then it cannot distinguish between a legitimate user and imposters \citep{baig2021security}.

\citet{balagani:2018} proposed a periodic keystroke dynamics-based privacy-preserving authentication scheme.
It is  similar to \citet{govindarajan:2013}, but it assumes the private comparison protocol of \citet{erkin:2009} in addition to the homomorphic DGK encryption algorithm of \citet{Damgaard:2008}.
This scheme has the same efficiency problems as Govindarajan et al.

\citet{wei:2020} proposed a privacy-preserving authentication scheme for touch dynamics using homomorphic encryption properties.
It is based on similarity scores between input and reference features using cosine similarity.
The authentication server performs a comparison between the encrypted reference template (provided during enrollment) and encrypted input template sampled during authentication.
The authentication server decrypts the similarity scores and compares them with a predefined threshold.




\section{\uppercase{Preliminaries}} \label{sec:prelim}

In this section we present some details on the Paillier cryptosystem and how it realizes its homomorphic properties.

\subsection{Briefly about the Paillier cryptosystem} \label{sec:Paillier_briefly}

Computations in the Paillier public key cryptosystem are conducted modulus~$n^2$, where $n=p q$, and $p$ and $q$ are large distinct primes of about the same size.
The public key is constituted by $(g,n)$, where $g = k n + 1$  and $k \geq 1$.
For convenience, let $g=n+1$.
The private key consists of~$(\lambda,n)$, where $\lambda = \lambda(n) = \lcm(p-1,q-1)$ is the Carmichael function.
As a side note, $\lambda(n)$ is a reduced version of the Euler totient function $\phi(n)=(p-1)(q-1)$, since $\lambda(n)$ divides $\phi(n)$.
Therefore, $\phi(n)$ can be used as the private key.

Encryption is conducted by means of the public key $(g,n)$ according to
$c=g^m r^n = (1 + m n) r^n \bmod{n^2}$,
where $r$ is a secret random integer selected by the sender.
We refer to $r^n$ as a Paillier encryption factor.
At decryption, this is eliminated by means of the private key~$(\lambda,n)$, since
$r{n \lambda} \equiv 1 \pmod{n^2}$.
Likewise, $r{n \phi(n)} \equiv 1 \pmod{n^2}$.

The decisional composite residuosity (DCR) assumption states that it is hard to decide whether $z$ is an $n$-residue modulo $n^2$, that is, whether there exists a number $r \in \mathbb{Z}_{n^2}^*$ so that $z = r^n \bmod {n^2}$.
This means that given a Paillier ciphertext $c$; if $r$ is an unknown random integer generated by the sender who computed $c$, then $r^n$ is similarly hard to determine.

\subsection{Homomorphic properties of the Paillier public key cryptosystem}


Consider the binomial expansion
$$
(1 + n)^x = \sum_{j=0}^{x} \binom{x}{j} = \sum_{j=0}^x k_j x^j  = 1 +  x n + \ldots + n^x
$$
where $k_j$, $0 \leq j \leq x$, are binomial coefficients.
Since all computations are conducted modulo~$n^2$, all terms having the factor $n^2$ become eliminated, and so
\begin{equation*}  
 H(x) = g^x \equiv  (1 + n)^x \equiv 1 + x n \pmod {n^2}
\end{equation*}
The additive homomorphic property is reflected by
\begin{equation*}
\begin{split}
H(x_1) H(x_2) & =  g^{x_1} g^{x_2} \equiv (1 + x_1 n)(1 + x_2 n)  \pmod {n^2} \\
      & \equiv 1 + (x_1 + x_2) n = H(x_1 + x_2) \pmod {n^2}
\end{split}
\end{equation*}
and
\begin{equation*}
\begin{split}
\frac{H(x_1)}{H(x_2)} &= \frac{g^{x_1}}{g^{x_2}} 
\equiv \frac{1 + x_1 n}{1 + x_2 n} \pmod {n^2}  \\
 & \equiv (1 + x_1 n)(1 - x_2 n) \pmod {n^2}  \\
   & \equiv 1 + (x_1 - x_2) n = H(x_1 - x_2)  \pmod {n^2}
\end{split}
\end{equation*}
and
\begin{equation*}
\begin{split}
H(x)^k &= (g^{x_1})^k = (1 + x n)^k  \pmod {n^2} \\
 &\equiv 1 + k x n = H(k x)  \pmod {n^2}
\end{split}
\end{equation*}


\section{\uppercase{The Wei et al. privacy-preserving authentication protocol}} \label{sec:scheme}

The \citeauthor{wei:2020} protocol involves two parties: A user $P_i$ and an authentication server (AS).
It consists of the following steps:

\vspace{2mm}

\noindent \emph{System initialization.}
The authentication server (AS) computes a Paillier key pair.
The public key consists of $(g,n)$, where $n=p q$ is a composite modulus of which $p$ and $q$ are two large and distinct primes, and $g = k n + 1$, for an integer $k \geq 1$.
For simplicity, let $g=n+1$.
The private key $(\lambda,n)$ is only known by AS.
In addition to the Paillier key pair of the AS, each user generates a secret key vector during encrollment.

\vspace{2mm}

\noindent \emph{User enrollment.}
The user enrollment process consists of three steps: User key generation, reference template sampling and generation, and encryption of the reference template.
A user $P_i$ samples behavior data for the reference template vector
$\vec{a}_i = (a_{i,1},\ldots, a_{i,t} )$. 
Then $P_i$ chooses two long-term secret encryption key vectors
$\vec{x}_i = (x_{i,1},\ldots, x_{i,t} )$ and $\vec{r}_i = (r_{i,1},\ldots, r_{i,t} )$, where each element is randomly chosen in $\mathbb{Z}_{n^2}^*$.
$\vec{x}_i$ is used for encryption of the reference template $\vec{a}_i $ and for encryption of behavior data in the subsequent authentication process.

$P_i$ encrypts each element in $\vec{a}_i$ according to
$$
\vec{c}_i  = (c_{i,j} = g^{a_{i,j} + x_{i,j} } r_{i,j}^n  \pmod {n^2}, \quad 1 \leq j \leq t)
$$
Note that the secret factors $r_{i,j}^n$, $1 \leq j \leq t$, are consistent with the Paillier encryption factor of the Paillier cryptosystem.
This means that in agreement with Paillier decryption, the AS, holding the private Paillier key, would be able to restore $(a_{i,j} + x_{i,j})$.
The purpose of $\vec{x}_i$ is therefore to protect $\vec{a}_i$ from the AS.

\vspace{2mm}

\noindent \emph{User authentication.}
When $P_i$ has collected a feature vector of sampled values $\vec{b}_i = (b_{i,1},\ldots, b_{i,t} )$, the authentication process is initiated.
It consists of the following three rounds:

\vspace{1mm}

\emph{Round 1.}
$P_i$ generates an ephemeral random vector $\vec{r}_i{}^* = (r_{i,1}^*,\ldots, r_{i,t}^* )$, where each element is selected in $\mathbb{Z}_{n^2}^*$, and encrypts each element in $\vec{b}_i = (b_{i,1},\ldots, b_{i,t} )$ according to
\begin{equation*} 
  c_{i,j}^* = (g^{b_{i,j}} r_{i,j}^*{}^n )^{x_{i,j}}
  = g^{b_{i,j} x_{i,j}} r_{i,j}^*{}^{n \, x_{i,j}} \pmod {n^2}
\end{equation*}
where $r_{i,j}^*{}^n$, $ 1 \leq j \leq t$, are consistent with a Paillier encryption factor having $n$ in the exponent.
$P_i$ sends the encrypted feature vector
$\vec{c}_i{}^* = (c_{i,1}^*,\ldots, c_{i,t}^* )$
to AS.

\vspace{1mm}

\emph{Round 2.}
The AS receives $\vec{c}_i{}^*$ and retrieves the encrypted enrollment vector $\vec{c}_i$ of $P_i$.
AS generates an epehemral random vector $\vec{r}_i{}' = (r_{i,1}',\ldots, r_{i,t}' )$, and blinds each element of the encrypted template vector $\vec{c}_i$ according to:
$$
 c_{i,j}' = c_{i,j}^{ r_{i,j}' } = g^{ (a_{i,j} + x_{i,j}) r_{i,j}' } r_{i,j}^{n \, r_{i,j}'} \pmod {n^2}, \quad 1 \leq j \leq t
$$
and sends the vector
$\vec{c}_i{}' = (c_{i,1}',\ldots, c_{i,t}' )$
to $P_i$.

\vspace{1mm}

\emph{Round 3.}
$P_i$ receives $\vec{c}_i{}'$ and computes
\begin{equation} \label{eq:round3}
  d_{i,j} = c'_{i,j}{}^{b_{i,j}}
    = g^{ (a_{i,j} + x_{i,j}) r_{i,j}' b_{i,j} } r_{i,j}^{n \, r_{i,j}' b_{i,j} } \pmod {n^2}
\end{equation}
for $1 \leq j \leq t$, and sends the vector
$\vec{d}_i = (d_{i,1},\ldots, d_{i,t})$
to AS.

\vspace{1mm}

\emph{Authentication decision.} The AS holds now $( \vec{c}_i^*, \vec{d}_i, \vec{r}_i' )$.
By means of $\lambda$, the AS inverts the elements in $\vec{r}_i'$ modulo $\phi(n^2) = n \lambda$, and then computes
\begin{equation*}
\begin{split}
  t_{i,j} &= \left( \frac{d_{i,j}}{ c_{i,j}^*{}^{ r_{i,j}'}  } \right)^{ (r_{i,j}')^{-1}}
  =   \frac{d_{i,j}^{ (r_{i,j}')^{-1}}}{ c_{i,j}^*} \pmod {n^2} \\
  &= \frac{\big( g^{ (a_{i,j} + x_{i,j}) r_{i,j}' b_{i,j} } r_{i,j}^{n \, r_{i,j}' b_{i,j} } \big)^{ (r_{i,j}')^{-1}} }{g^{b_{i,j} x_{i,j}} r_{i,j}^*{}^{n \, x_{i,j}} }  \pmod {n^2} \\
 & = \frac{ g^{ a_{i,j}  b_{i,j} + x_{i,j}  b_{i,j} } r_{i,j}^{n \, b_{i,j}} }{g^{b_{i,j} x_{i,j}} r_{i,j}^*{}^{n \, x_{i,j}} }
   = \frac{ g^{ a_{i,j}  b_{i,j}} g^{ x_{i,j}  b_{i,j} } }{g^{b_{i,j} x_{i,j}}  } R_{i,j}^n  \pmod {n^2} \\
 &= g^{ a_{i,j} b_{i,j} } R_{i,j}^n   \pmod {n^2}, \quad 1 \leq j \leq t
  \end{split}
\end{equation*}
where $R_{i,j}^n =  r_{i,j}^{n \, r_{i,j}' b_{i,j} }  / r_{i,j}^*{}^{n \, x_{i,j}} $
is consistent with a Paillier encryption factor.
AS aggregates
$$
T = \prod_{j=1}^{t} t_{i,j} = g^{ \vec{a}_{i} \vec{b}_{i} } R_i^n  \pmod{n^2}
$$
where $R_i^n$ is an aggregated Paillier encryption factors.
AS decrypts $T$ in agreement with the Paillier cryptosystem, using its private key $\lambda$, to obtain the vector product
$ T' =  \vec{a}_{i} \vec{b}_{i} $.
Let $T_S$ be a predetermined threshold.
If $T' \geq T_S$ then $P_i$ is considered authentic, otherwise the authentication fails.

\vspace{2mm}

The primary purpose of the Paillier encryption factor is to blind the plaintext so that it becomes unintelligible to anyone not holding the private key.
The designated recipient of the ciphertext, holding the private key, decrypts the ciphertext which removes the secret encryption factor (see Section~\ref{sec:Paillier_briefly}), and the plaintext is restored.



\section{\uppercase{Cryptanalysis}}

In this section we show that the Wei et al. protocol is insecure with regard to passive and active attacks.
In any case, it is insecure due to its homomorphic property.

\subsection{Honest-but-curious authentication server attack}

An honest-but-curious adversary is an adversary who does not deviate from the defined protocol by modifying or computing messages in ways that are not in agreement with the protocol.
It will rather attempt to learn all possible information from the received messages and other legitimate information it may hold, such as previous messages and public keys.
As shown as follows, an honest-but-curious authentication server is able to obtain not only the plaintext timeseries vector $\vec{b}_i$, but also the secret user key vector $\vec{x}_i$ and the plaintext feature template vector $\vec{a}_i$.

During enrollment, AS receives $\vec{c}_i$ from $P_i$.
Decrypting each element in $\vec{c}_i$ in agreement with the Paillier decryption algorithm removes the encryption factor $r_{i,j}^{n}$:
\begin{equation*}
\begin{split}
 C_{i,j} &= c_{i,j}^\lambda = g^{(a_{i,j} + x_{i,j}) \lambda} r_{i,j}^{n \, \lambda} \pmod {n^2} \\
         &= g^{(a_{i,j} + x_{i,j}) \lambda} = 1+ (a_{i,j} + x_{i,j}) n \lambda  \pmod {n^2}
\end{split}
\end{equation*}
and restores the sum of the elements of the enrollment vector $\vec{a}_i$ and key vector $\vec{x}_i$:
\begin{equation} \label{eq:c}
\begin{split}
L(C_{i,j}) &=  \frac{C_{i,j} -1}{n \lambda}  \\
  &= \frac{1+ (a_{i,j} + x_{i,j}) n \lambda - 1}{n \lambda } = a_{i,j} + x_{i,j}
\end{split}
\end{equation}

\noindent In Round~3 the AS receives encrypted sampled vector~$\vec{d}_i$.
Decrypting each element in agreement with Paillier decryption removes the Paillier encryption factor $r_{i,j}{}^{n \, r_{i,j}' b_{i,j}}$ from~$d_{i,j}$:
\begin{equation*}
\begin{split}
 D_{i,j} &= d_{i,j}^{ (r_{i,j}')^{-1} \lambda } 
 = g^{ (a_{i,j} + x_{i,j}) b_{i,j} r_{i,j}' (r_{i,j}')^{-1} \lambda} r_{i,j}^{n r_{i,j} b_{i,j} (r_{i,j}')^{-1} \lambda  } \\
  &= g^{ (a_{i,j} + x_{i,j}) b_{i,j} \lambda}  \\
  &= 1+ (a_{i,j} + x_{i,j}) b_{i,j}\lambda n   \pmod {n^2} , \quad 1 \leq j \leq t
 \end{split}
\end{equation*}
and eventually
\begin{equation} \label{eq:d}
 \begin{split}
 L(D_{i,j}) &=  \frac{D_{i,j} -1 }{\lambda n}
  = \frac{1+ (a_{i,j} + x_{i,j}) b_{i,j}\lambda n - 1}{\lambda n} \\
  &= (a_{i,j} + x_{i,j}) b_{i,j}
 \end{split}
\end{equation}
Dividing Eqs.~(\ref{eq:d}) and (\ref{eq:c}) reveals the sampled plaintext vector $\vec{b}_i$, in which
$$
 b_{i,j} = \frac{(a_{i,j} + x_{i,j}) b_{i,j}}{a_{i,j} + x_{i,j}}, \quad 1 \leq i \leq t
$$

The secret key vector $\vec{x}_i$ of $P_i$ is restored as follows.
In Round 1, $P_i$ sends the ciphertext $c_{i,j}^*$.
Decryption yields
\begin{equation*}
\begin{split}
 C_{i,j}^* &= c_{i,j}^*{}^\lambda
 = g^{b_{i,j} x_{i,j} \lambda} r_{i,j}^*{}^{n \, x_{i,j} \lambda}  \\
 &= g^{b_{i,j} x_{i,j} \lambda}
 = 1 + b_{i,j} x_{i,j} \lambda n \pmod {n^2}
 \end{split}
\end{equation*}
and then
\begin{equation} \label{eq:c*}
 L(C_{i,j}^*) = \frac{ C_{i,j}^* -1}{\lambda n}
  = \frac{1 + b_{i,j} x_{i,j} \lambda n -1}{\lambda n} = b_{i,j} x_{i,j}
\end{equation}
AS can now restore the secret key vector $\vec{x}_i$ by means of Eqs.~(\ref{eq:c*}) and~(\ref{eq:c}):
$$
x_{i,j} = \frac{b_{i,j} x_{i,j}}{b_{i,j} }, \quad 1 \leq i \leq t
$$
By means of $x_{i,j}$ and Eq.~(\ref{eq:c}) the plaintext template vector $\vec{a}_i$ is recovered:
$$
a_{i,j} = (a_{i,j} + x_{i,j}) - x_{i,j}, \quad 1 \leq i \leq t
$$


\subsection{Active adversary attack}

In the following, we show that the Wei et al. scheme is insecure to active attacks that include an active eavesdropper $\mathcal{A}$ who is capable of modifying the communication between AS and the user $P_i$.
The attack can be conducted so that neither $P_i$ or AS will be aware of the ongoing attack.

In Round~2, $\mathcal{A}$  computes a random vector
$$
\vec{c}_i{}'' = (c_{i,j}'' = 1 + n \,r_{i,j}'' \bmod{n^2}, \quad 1 \leq j \leq t)
$$
where $r_{i,j}''$, $1 \leq j \leq t$, are random integers.
$\mathcal{A}$ replaces the legitimate vector $\vec{c}_i{}'$ from $P_i$ with $\vec{c}_i{}''$.
Note that this will not be noticeable since the Wei et al. scheme does not provide a means to verify the authenticity of $\vec{c}_i{}''$.

In Round~3, $P_i$ receives $\vec{c}_i{}''$, and computes
$$
d_{i,j} = c_{i,j}''{}^{b_{i,j}} = 1 + n \,r_{i,j}'' b_{i,j}  \pmod {n^2}, \quad 1 \leq i \leq t
$$
in agreement with Eq.~(\ref{eq:round3}).
$\mathcal{A}$ receives $\vec{d}_i$, and restores
$$
b_{i,j} = \frac{d_{i,j} -1}{r_{i,j}'' n}
        = \frac{1 + b_{i,j} r_{i,j}'' n -1}{r_{i,j}'' n }, \quad 1 \leq i \leq t
$$

In order to keep the AS unaware of the attack, $\mathcal{A}$ conducts Round~2 on behalf of $P_i$ by computing and forwarding
$d_{i,j} = c'_{i,j}{}^{b_{i,j}}$
to AS, whom receives a correct and legitimate computation.

Note that $\mathcal{A}$ is unable to eliminate the intrinsic Paillier encryption factors of $\vec{c}_i'$ due to not knowing the private key $\lambda$.
The plaintext template vector $\vec{a}_i$, supplied during enrollment and the secret key vector $\vec{x}_i$ used during enrollment and Round~1 remain as such protected.


\section{\uppercase{Conclusion}}

Continuous authentication has been proposed as an alternative to password-based authentication for computer devices that are handled directly by humans, such as smart phones.
However, continuous authentication has some privacy issues as certain user features and actions are revealed to the authentication server.
In this paper, we have considered a privacy-preserving protocol for behavioral authentication proposed by Wei et al. in 2021.
We show that their scheme is insecure with regard to an honest-but-curious server and an active eavesdropper.
We present two attacks, in which the first enables the authentication server to obtain behavioral template plaintext data, the authentication plaintext data, and the user's secret encryption key plaintext.
The second attack enables an active eavesdropper to obtain authentication plaintext data.

The foundational problem is the way that encryption is conducted during enrollment by means of user key vector, and that this does not provide any real protection regarding the authentication server, who can remove the Paillier encryption factor, and then algebraically compromise data and user key vectors.
Another foundational problem regarding active adversaries is that there is no way to detect message modification. This is something that would be detected by most cryptographic authentication protocols, where authentication is based on zero-knowledge proofs of some cryptographic keys, but in behavioral authentication schemes the authentication concept is consequently not based on keys, although cryptographic methods have been proposed for resolving the privacy issues of behavioral authentication schemes.
As such, we do not see an immediate way to how to fix the problem.

\section*{\uppercase{Acknowledgement}}
{This work is part of the Privacy Matters (PriMa) project. The PriMa project has received funding from European Union’s Horizon 2020 research and innovation programme under the Marie Skłodowska-Curie grant agreement No. 860315.}

\bibliographystyle{apalike}
\bibliography{report}  

\end{document}